\documentclass[12pt]{article}
\usepackage{latexsym,cite}
\newcommand{\eq}{\begin{equation}}
\newcommand{\en}{\end{equation}}
\newcommand{\eqa}{\begin{eqnarray}}
\newcommand{\ena}{\end{eqnarray}}

\newcommand{\PRL}[1]{Phys.\ Rev.\ Lett.\ {\bf #1}}

\begin{document}

\title{Nonanalyticity of the Callan-Symanzik $\beta$-function of two-dimensional O($N$) models}
\author{
  \\
  {\small Pasquale Calabrese}              \\[-0.2cm]
  {\small\it Dipartimento di Fisica, Universit\`a degli Studi di Pisa,
    I-56127 Pisa, ITALIA}        \\[-0.2cm]
  \\[-0.1cm]  
  {\small Michele Caselle}              \\[-0.2cm]
  {\small\it Dipartimento di Fisica and INFN -- Sezione di Torino}    \\[-0.2cm]
  {\small\it Universit\`a degli Studi di Torino}        \\[-0.2cm]
  {\small\it I-10125 Torino, ITALIA}          \\[-0.2cm]
  {\small caselle@to.infn.it}          \\[-0.2cm]
  \\[-0.1cm]  
  {\small Alessio Celi}              \\[-0.2cm]
  {\small\it Dipartimento di Fisica, Universit\`a degli Studi di Pisa,
   I-56127 Pisa, ITALIA}        \\[-0.2cm]
  \\[-0.1cm]  
  {\small Andrea Pelissetto}              \\[-0.2cm]
  {\small\it Dipartimento di Fisica and INFN -- Sezione di Roma I}    \\[-0.2cm]
  {\small\it Universit\`a degli Studi di Roma ``La Sapienza"}        \\[-0.2cm]
  {\small\it I-00185 Roma, ITALIA}          \\[-0.2cm]
  {\small Andrea.Pelissetto@roma1.infn.it}          \\[-0.2cm]
  \\[-0.1cm]  \and
  {\small Ettore Vicari}              \\[-0.2cm]
  {\small\it Dipartimento di Fisica and INFN -- Sezione di Pisa}    \\[-0.2cm]
  {\small\it Universit\`a degli Studi di Pisa}        \\[-0.2cm]
  {\small\it I-56127 Pisa, ITALIA}          \\[-0.2cm]
  {\small vicari@df.unipi.it}          \\[-0.2cm]
  {\protect\makebox[5in]{\quad}}  % To force authors' names to be written
                                  %   vertically, one above another.
                                  % (\author seems to put them side-by-side
                                  %   if there is room.)
  \\
}
\vspace{0.5cm}
\date{September 5, 2000}
%\date{\today}

\maketitle
\thispagestyle{empty}   % Suppress page number on front page.

\vspace{0.2cm}

\begin{abstract}
We discuss the analytic properties of the Callan-Symanzik $\beta$-function
$\beta(g)$ associated with the zero-momentum four-point 
coupling $g$ in the two-dimensional $\phi^4$ model with $O(N)$
symmetry. Using renormalization-group arguments, 
we derive the asymptotic behavior of  $\beta(g)$ at the fixed point 
$g^*$. We argue that $\beta'(g) = \beta'(g^*) + O(|g-g^*|^{1/7})$ 
for $N=1$ and $\beta'(g) = \beta'(g^*) + O(1/\log |g-g^*|)$ for 
$N\ge 3$. Our claim is supported by an explicit calculation in the 
Ising lattice model and by a $1/N$ calculation for the two-dimensional $\phi^4$
theory. We discuss how these nonanalytic corrections may give rise to a slow 
convergence of the perturbative expansion in powers of $g$. 
\end{abstract}

\clearpage

\newcommand{\be}{\begin{equation}}
\newcommand{\ee}{\end{equation}}
\newcommand{\bea}{\begin{eqnarray}}
\newcommand{\eea}{\end{eqnarray}}
\newcommand{\<}{\langle}
\renewcommand{\>}{\rangle}

%\ltapprox and \gtapprox produce > and < signs with twiddle underneath
\def\spose#1{\hbox to 0pt{#1\hss}}
\def\ltapprox{\mathrel{\spose{\lower 3pt\hbox{$\mathchar"218$}}
 \raise 2.0pt\hbox{$\mathchar"13C$}}}
\def\gtapprox{\mathrel{\spose{\lower 3pt\hbox{$\mathchar"218$}}
 \raise 2.0pt\hbox{$\mathchar"13E$}}}

\def\bsigma{\mbox{\protect\boldmath $\sigma$}}
\def\btau{\mbox{\protect\boldmath $\tau$}}
\def\bphi{\mbox{\protect\boldmath $\phi$}}
\def\bz{\mbox{\protect\boldmath $z$}}
\def\bw{\mbox{\protect\boldmath $w$}}
\def\hatp{\hat p}
\def\hatl{\hat l}
\def\smfrac#1#2{{\textstyle\frac{#1}{#2}}}
\def\case#1#2{{\textstyle\frac{#1}{#2}}}

\def\msbar{ {\overline{\hbox{\scriptsize MS}}} }
\def\normalmsbar{ {\overline{\hbox{\normalsize MS}}} }

\newcommand{\R}{\hbox{{\rm I}\kern-.2em\hbox{\rm R}}}
\newcommand{\N}{\hbox{{\rm I}\kern-.2em\hbox{\rm N}}}

\newcommand{\reff}[1]{(\ref{#1})}

\section{Introduction}

Renormalization-group theory 
is a very important tool for the understanding of the critical behavior 
of statistical models in the neighbourhood of the critical point.
We consider models with an $N$-vector real order parameter and $O(N)$
symmetry. Because of universality, quantitative predictions can be obtained by 
studying any theory belonging to the 
same universality class. For the models we are dealing with here, 
we may consider the Ginzburg-Landau Hamiltonian
\begin{equation}
{\cal H} = \int d^dx\left[ {1\over 2}\left( \partial_\mu \vec{\phi}\,\right)^2 
+ {1\over 2} r \vec{\phi}^{\,2} + {1\over 4!}g_0(\vec{\phi}^{\,2})^2\right],
\label{Hphi4}
\end{equation}
where $\vec{\phi}$ is an $N$-component real field.
This Hamiltonian describes many interesting systems at criticality. 
The liquid-vapour transition in fluids 
and the infinite-length properties of polymers in dilute solutions  
correspond to the $N=1$ (Ising) and $N=0$ model respectively;
the ${}^4$He superfluid phase transition is in the same universality class of 
the three-dimensional two-component theory ($XY$ model), while
the Hamiltonian \reff{Hphi4} with $N=3$ describes
isotropic ferromagnetic materials.
Three-dimensional $N$-vector systems and two-dimensional systems with $N<2$
have a conventional critical behavior: Thermodynamic quantities 
have power-law singularities near the critical point.
On the other hand, in two dimensions the $XY$ model shows a 
Kosterlitz-Thouless transition, while
for $N\geq 3$ no finite-temperature transition exists: The correlation
length diverges only for $T\to 0$. For $N\geq 3$ 
the theory is asymptotically free 
with a critical behaviour described by the perturbative 
renormalization group applied to the nonlinear $\sigma$-model.

Precise estimates of the critical parameters in the symmetric phase 
can be obtained using several different methods. One of them, which 
provides in many cases very precise results, 
relies on a perturbative expansion in powers of the 
zero-momentum four-point renormalized coupling $g$
performed at fixed dimension $d$~\cite{Parisi-80}.
The theory is renormalized by introducing
a set of zero-momentum conditions for the (one-particle irreducible) 
two-point and four-point correlation functions:
\begin{eqnarray}
&&\Gamma^{(2)}(p)_{\alpha\beta} = \delta_{\alpha\beta}\;
Z_G^{-1} \left[ m^2+p^2+O(p^4)\right],
\label{p2}  \\
&&\Gamma^{(4)}(0,0,0,0)_{\alpha\beta\gamma\delta} =
Z_G^{-2} \,m^{4-d}\, g \, {1\over 3}
\left( \delta_{\alpha\beta}\delta_{\gamma\delta}
+\delta_{\alpha\gamma}\delta_{\beta\delta}
+\delta_{\alpha\delta}\delta_{\beta\gamma}\right).
\label{p4}
\end{eqnarray}
For $m\rightarrow 0$, the coupling $g$ is driven toward an
infrared-stable zero $g^*$ of the corresponding Callan-Symanzik
$\beta$-function
\begin{equation}
\beta(g)\equiv m {\partial g \over \partial m}.
\end{equation}
The derivative of the $\beta$-function at $g^*$, $\beta'(g^*)$, is related 
to the leading nonanalytic correction-to-scaling exponent. 
Usually---but we shall argue here that this may not always be the case---the
leading nonanalytic corrections are determined by the critical dimension
$\omega_1$ of the leading irrelevant operator:
in this case, we have $\beta'(g^*)=\omega_1$.
At present, $\beta(g)$ has been computed to 
six loops in three dimensions~\cite{BNGM-77}
and to five loops in two dimensions~\cite{OS-00}.

Perturbative expansions in powers of $g$ are asymptotic. 
In order to obtain estimates of universal critical quantities, 
it is essential to resum the perturbative series. This can be done
by exploiting their Borel summability
and the knowledge of their large-order behavior
(see e.g. \cite{ZJ-book} and references therein).
The large-order behavior of the series $S(g) = \sum s_k g^k$
is related to the singularity $g_b$ of the Borel transform $B(g)$
that is closest to the origin. For large $k$,
\begin{equation}
s_k \sim k! \,(-a)^{k}\, k^b \,\left[ 1 + O(k^{-1})\right] \qquad\qquad
{\rm with}\qquad a = - 1/g_b.
\label{lobh}
\end{equation}
The value of $g_b$ can be obtained by means of  
a steepest-descent calculation~\cite{Lipatov-77,BLZ-77}.
It depends only on the Hamiltonian,
while the exponent $b$ depends on which Green's function is considered.
If the perturbative expansion is Borel summable, then $g_b$ is negative.
Since the Borel transform is singular for $g=g_b$, its expansion in powers 
of $g$ converges only for $|g|< |g_b|$.
An analytic extension can be obtained by a conformal 
mapping~\cite{LZ-77}, such as
\begin{equation}
y(g) = {\sqrt{1 - g/g_b} - 1\over \sqrt{1 - g/g_b} + 1 }.
\label{cfmap}
\end{equation}
The Borel transform becomes an expansion in powers of $y(g)$
that converges for all positive values of $g$, provided that all 
singularities of the Borel transform are on the real negative 
axis~\cite{LZ-77}.  Therefore, the
use of the Borel transform and of the conformal mapping (\ref{cfmap})
transforms the original asymptotic series into a convergent
expansion. 
Any universal quantity, such as the critical exponents,
is estimated by resumming the corresponding perturbative
series and by evaluating the resummed function of $g$ at 
the fixed-point value $g^*$.

The critical value $g^*$ of the renormalized coupling is a universal quantity.
Therefore, it 
can also be obtained by considering any statistical (lattice) model
belonging to the corresponding universality class. 
Then
\begin{equation}
g^* = \lim_{t \to 0} g(t) \equiv \lim_{t\to 0} \left[
 - {3N\over N+2}\ {\chi_4 \over \chi^2\xi^d } \right],
\label{defgl}
\end{equation}
where  $t\equiv T/T_c - 1$,
$\chi$ is the magnetic susceptibility, $\xi$ the second-moment
correlation length, and 
$\chi_4$ the zero-momentum four-point connected
correlation function. Using Eq.~(\ref{defgl}), one can obtain 
an independent estimate of $g^*$.

An important issue in the field-theoretical (FT) approach concerns the
analytic properties of $\beta(g)$. General renormalization-group
arguments~\cite{Parisi-80,Nickel-82,Nickel-91} (see also 
\cite{Sokal-94,BB-97})
and explicit calculations to next-to-leading order 
in the framework of the $1/N$ expansion~\cite{PV-98,CPRV-96} 
show that $\beta(g)$ is not analytic at $g=g^*$.
This fact 
may cause a slow convergence of the resummations of the perturbative 
series to the correct fixed-point value. The reason is that
this resummation method approximates the $\beta$-function in the interval 
$[0,g^*]$ with a sum of analytic functions. Since, for $g=g^*$, the 
$\beta$-function is not analytic, the convergence at the endpoint
of the interval is slow. 
This may also lead to an underestimate of the uncertainty that is 
usually derived from stability criteria.  
In spite of these problems, in three dimensions, 
FT results are in good agreement\footnote{ 
Small discrepancies 
are only observed for $N=0$ and $N=1$. For instance,
we may compare the estimates of $g^*$ and $\omega_1$ 
obtained using the fixed-dimension FT approach
with  the apparently best estimates obtained from the analysis 
of high-temperature (HT) expansions and from Monte Carlo (MC) simulations for 
lattice models in the same universality class.
For $N=1$, the analysis of the fixed-dimension FT expansion
gives $g^*=1.411(4)$ and $\omega_1=0.799(11)$~\cite{GZ-98}, 
to be compared with the lattice results $g^*=1.402(2)$~\cite{CPRV-99} (HT) and
$\omega_1=0.845(10)$~\cite{Hasenbusch-99} (MC).
The results are in better agreement for $N=2$: 
the analysis of the fixed-dimension
$g$-expansion leads to $g^*=1.403(3)$ and $\omega_1=0.789(11)$~\cite{GZ-98},
to be compared with 
$g^*=1.396(4)$~\cite{CPRV-00} (HT) and $\omega_1=0.79(2)$~\cite{HT-99} (MC).  }
%% fine footnote
with the estimates obtained in other approaches
\cite{PV-98,GZ-98,CPRV-99,Hasenbusch-99,HT-99,CPRV-00,Zinn-Justin-00},
showing that 
the above-mentioned nonanalyticity causes only very small effects
that are negligible in most cases.
Using general renormalization-group arguments, 
for three-dimensional models one expects \cite{Nickel-82} 
\be
\beta(g) = -\beta'(g^*) (g^* - g) \left[
  1 + a_1 (g^* - g)^p + a_2 (g^* - g) + \cdots \right],
\label{bgexp}
\ee
where $\beta'(g^*)=\omega_1$
and $p$ is a noninteger exponent that is equal to the smallest of the 
following exponent combinations:
$p=\omega_2/\omega_1-1$ 
where $\omega_2$ is the scaling dimension of the next-to-leading 
irrelevant operator, $p=1/\Delta$ where $\Delta = \omega_1 \nu$,
and $p=\gamma/\Delta-1$.
Notice the last exponent that was neglected in \cite{Nickel-82,PV-98}
and that is due to a subleading correction in 
$g(t)$ proportional to $t^\gamma$. Such a term is related 
to the presence of an analytic background in the free energy.
For small values of $N$, we have $\Delta\equiv \omega_1 \nu \approx 1/2$,
$\omega_2/\omega_1 \approx 2$~\cite{GRNR-76}, and $\gamma/\Delta> 2$, 
so that $p  = \omega_2/\omega_1 - 1\approx 1$. 
In this case the leading nonanalytic term is practically
undistinguishable from the 
analytic one, and therefore, one expects only small systematic deviations.
For increasing values of $N$, $p$ decreases, but at the same time $a_1\to 0$.
Thus, also in this case we expect the nonanalytic terms to give rise to small
systematic deviations.

The situation worsens in the two-dimensional case which we
consider in  this report. As a matter of fact, at variance with the 
three-dimensional case, two-dimensional FT estimates 
are much more imprecise \cite{OS-00}. We shall argue here that 
the large observed deviations are caused by the nonanalyticity of 
the renormalization-group functions at $g^*$. 
In order to support this argument, we shall compute the behaviour 
of the $\beta$-function for $g\to g^*$ in two cases
in which exact results can be obtained exploiting  
different techniques. 

First, we shall address the $N=1$ case 
(i.e. the Ising universality
class) in which conformal field theory (CFT) techniques allow the determination 
of the whole spectrum of relevant and irrelevant operators of the theory.
We shall first show that CFT predicts $\omega_1 = 2$ for the 
renormalization-group dimension of the leading irrelevant operator, 
excluding $\omega_1 = 4/3$, as it has been claimed sometimes. 
Then, we will consider the lattice Ising model
and we will show that Eq. \reff{bgexp} holds with 
$\beta'(g^*)=\gamma/\nu=7/4$ and $p=1/7$. Notice that in this case 
$\beta'(g^*)\not=\omega_1 = 2$. \footnote{
This is due to the fact that the correction-to-scaling term with the 
smallest exponent appearing in $g(t)$ is $t^{\gamma}$ and not 
$t^{\omega_1\nu}$. A correction term proportional to $t^\gamma$ in $g(t)$ 
is due to the presence of an analytic background in the free energy.}
We will then argue that this is the generic behaviour one should expect 
for models in the Ising universality class. At variance with the 
three-dimensional case, here $p$ is very small and thus it may be 
responsible for large systematic deviations
in the resummation of the perturbative series. 
In App. B we study some simple Borel-summable asymptotic series behaving 
as \reff{bgexp} with $p=1/7$. We apply the resummation method 
describe above, finding very poor estimates of $\beta'(g^*)$ 
with largely underestimated error bars.

Second, we shall study  the multicomponent 
$\phi^4$ theory with $N\geq 3$. Since the model is asymptotically free, 
we can predict $\omega_1 = 2$ and we can show that logarithmic 
corrections should be expected at the critical point. A large-$N$
calculation confirms the theoretical predictions.

\section{$N=1$ $\phi^4$ theory in $d=2$}

Let us first consider the Ising case, i.e. the case in which the field
$\phi(x)$ in the $\phi^4$ Hamiltonian is a one-component
real field. In \cite{LZ-77} the four-loop series of $\beta(g)$
is analyzed using the resummation procedure presented in the introduction:
they obtain
$g^* = 15.5(8)$ and $\beta'(g^*) = 1.3(2)$.  Reference \cite{OS-00} 
computes the five-loop contribution and presents an analysis of the 
extended series using a Pad\'e-Borel resummation: they
obtain\footnote{We applied the Le Guillou--Zinn-Justin 
resummation method~\cite{LZ-77,ZJ-book}, 
using the conformal mapping (\ref{cfmap}),
to the five-loop series
of \cite{OS-00}: we obtained substantially equivalent results.}
$g^*=15.39(25)$ and $\beta'(g^*) = 1.31(3)$.
These results for $g^*$ do not agree with the very precise estimates 
obtained by a transfer-matrix analysis of the 
standard square-lattice Ising model~\cite{CHPV-00},
$g^*=14.69735(3)$, and 
by exploiting the form-factor bootstrap approach~\cite{BNNPSW-00},
$g^* = 14.6975(1)$ (see also \cite{BC-96,ZLF-96,PV-98} for high-temperature 
results).
The result for $\beta'(g^*)$ has been interpreted~\cite{ZJ-book,OS-00}
as an indication in favor of the 
exact result $\beta'(g^*)=4/3$ that would imply the existence of an irrelevant 
operator with $\omega_1=4/3$.
However, the corresponding scaling corrections do not appear in the standard
lattice Ising model in which, thanks
to the known exact results (see e.g. \cite{MW-73,WMTB-76,McCoy-95}),
a detailed analysis of the leading correction terms is possible.
In principle, this fact does not imply that the interpretation
of~\cite{ZJ-book,OS-00} is wrong, since it could  be simply
explained by the absence of the corresponding irrelevant operator
in the lattice Ising model, which is only one of the possible 
realizations of the $\phi^4$ universality class. 
However, we shall show below that this is
not the case and that no subleading operator with $\omega_1=4/3$ exists in
any unitary model belonging to the
Ising universality class. In particular, it does not exist in the $N=1$ 
$\phi^4$ theory.

Let us briefly comment on this last point.
The $\omega_1=4/3$ interpretation was supported by the fact that
an operator with renormalization-group dimension
$\omega_1=4/3$ exists in a particular  nonunitary extension of the
Ising universality class which is conjectured to describe Ising percolation.
However, such an operator can only exist in nonunitary theories, and as a
consequence, it cannot be observed in the unitary $\phi^4$ theory.
We shall argue in this paper that the estimate of $\omega_1$ obtained 
in the framework of the perturbative expansion at fixed dimension is
strongly affected by nonanalytic corrections in the $\beta$-function.
The fact that one obtains $\beta'(g^*)\simeq 4/3$ is only a coincidence and 
is not related to the presence of the nonunitary operator with $\omega_1=4/3$ 
mentioned above. In order to
clarify the issue we have added in App.~\ref{app1}  a discussion on the
nonunitary extension of the Ising universality class and its relation with
the Ising percolation problem.

The only ingredients that are needed to extend the Ising result---the 
absence of an exponent $\omega_1=4/3$--- to the most
general unitary model in the $N=1$ $\phi^4$ universality class  
are Wilson's renormalization group and some basic results of CFT.

In Wilson's approach, we can rewrite ${\cal H}$ as 
\be
{\cal H} = {\cal H}^* + \sum_{\{ {\cal O} \} } u_{\cal O} (m) {\cal O},
\label{Hoffcritica}
\ee
where ${\cal H}^*$ is the fixed-point Hamiltonian, 
$\{{\cal O}\}$ a complete set of operators, and $u_{\cal O} (m)$ 
the corresponding
nonlinear scaling fields depending on the inverse correlation length
$m$. Then, we observe 
that the $\phi^4$ theory is unitary. This can be proved to all orders 
of perturbation theory. It can also be proved nonperturbatively 
by considering the lattice regularization of the model (\ref{Hphi4}). 
Indeed, the lattice theory corresponding to (\ref{Hphi4})---and, of course,
also the standard Ising model which is a particular limit of the 
lattice $\phi^4$ theory---with
nearest-neighbor couplings is exactly reflection positive, a property 
that guarantees the unitarity of the Minkowski theory. 
At the critical point the
theory becomes conformally invariant. Now the main point is that in the
framework of CFT there exists a complete classification of
all possible $Z_2$ symmetric unitary theories \cite{FQS-84,BPZ-84}.
Moreover, their
operator content is exactly known. 
This means that all dimensions of the operators ${\cal O}$
that may appear in Eq. (\ref{Hoffcritica}) giving rise to a
unitary theory are exactly known.\footnote
{Let us stress that our argument is by no means original. It was, for instance,
already present in~\cite{BF-85} that appeared right after the classification of
unitary CFT's. In this respect, our main new contribution 
is the exact calculation of $p$ and the use of this result (discussed in
detail in App. \ref{app2}) 
to show the relevance of nonanalytic corrections in the
FT Callan-Symanzik $\beta$-function.}
In particular, no operator with dimension $\omega_1=4/3$ exists.

According to the CFT analysis \cite{CH-00,CHPV-00}, the
leading irrelevant operator is 
$T\bar T$, where $T$ denotes the energy-momentum tensor,
which is expected to give rise to corrections of order $t^2$,
$t$ being the reduced temperature.
On the square lattice---but not in a rotationally-invariant model or 
on lattices with different rotational symmetry, for instance, on the 
triangular lattice---one must also consider a second operator,
${\cal T} = T^2+\bar T^2$, which is degenerate with the first one. 
While $T\bar T$ is rotationally invariant, $\cal T$
breaks rotational invariance and has only 
the reduced symmetry of the square lattice.
Since correlation function of $\cal T$ with rotationally 
invariant operators vanish, such operator should not contribute 
at order $t^2$ to observables that are rotationally invariant, but
only at order $t^4$ (indeed, $\< {\cal T}_x {\cal T}_y {\cal O}\>$
does not vanish even if ${\cal O}$ is rotationally invariant).
Of course, ${\cal T}$ should contribute to order $t^2$ to observables
that have an angular dependence (an explicit example will be given below).

In the last years there has been extensive work trying to understand the 
origin of the subleading corrections in the lattice Ising model. 
The unexpected result is the fact that no correction-to-scaling term 
due to $T\bar T$ has been observed. Let us review the evidence for this fact:
\begin{enumerate}
\item The analysis of the 
susceptibility~\cite{AF-83,Nickel-99,Guttmann-2000} for $h=0$
indicates that the corrections of order $t$, $t^2$, $t^3$ 
can be interpreted as purely analytic ones.
\item The analysis of the free energy on the critical isotherm 
as a function of $h$ \cite{CH-00} does not find any evidence of 
correction-to-scaling terms that can be associated to $T\bar{T}$. 
\item The analysis of the free energy, correlation length and 
susceptibility at the critical point in a finite box 
\cite{Queiroz-00,Hasenbusch-2000}
shows the presence of corrections with $\omega_1 = 2$. 
These corrections however appear to be due to $\cal T$ only. 
Indeed, they are not present on the triangular and honeycomb lattices
\cite{Queiroz-00}---on these lattices $\cal T$ cannot contribute and the 
first expected correction has $\omega_1 = 4$--- and moreover, 
the dependence of these corrections on the shape of the box is 
consistent with the behaviour expected for a spin-four 
operator as $\cal T$ is \cite{Hasenbusch-2000}.
\end{enumerate}
Here, we want to add further evidence for the absence of $T\bar{T}$ by 
considering the observables characterizing the large-distance 
behaviour of the two-point function on a square lattice. 
Indeed, for $x\to\infty$ we can write \cite{CW-67}
\be
\langle \sigma_0\sigma_x\rangle=\, 
Z(\beta) \int {d^2p\over (2\pi)^2} 
   {e^{ip\cdot x}\over \hat{p}^2 + M(\beta)^2},
\label{2ptlargex}
\ee
where $\hat{p}^2 = 4 \sum_\mu \sin^2(p_\mu/2)$ and the integration is 
extended over the first Brillouin zone. The quantities $Z(\beta)$ and 
$M(\beta)$ are known exactly \cite{CW-67}. For 
$t\equiv 1 - \beta/\beta_c\to 0$, we can write
\begin{eqnarray}
Z(\beta) &=& \left(128 \sqrt{2}\beta_c\right)^{1/4} u_t^{1/4} v_h^2 
   \left[1 + O(u_t^4)\right], \\
M(\beta)^2 &=& 16 \beta_c^2 u_t^2 \left[1 + \beta_c^2 u_t^2 + O(u_t^4)\right],
\end{eqnarray}
where $u_t$ is the nonlinear scaling field associated with the reduced 
temperature at zero magnetic field $h$ and $v_h$ is related to the 
nonlinear scaling field $u_h$ associated with 
$h$ by $u_h = h v_h + O(h^3)$. Explicitly \cite{SS-99,Nickel-99,CHPV-00} 
\begin{eqnarray}
u_t &=& t \left( 1 +  {\beta_c\over \sqrt{2}} t + 
                  {7\beta_c^2 \over 6} t^2 + {17\beta_c^3\over 6\sqrt{2}}
                  t^3 + O(t^4) \right), 
\label{utdef}\\
v_h &=& 1 + {\beta_c\over \sqrt{2}} t + 
      {23\beta_c^2 \over 16} t^2 + {191\beta_c^3\over 48\sqrt{2}}
                  t^3 + O(t^4) .
\label{vhdef}
\end{eqnarray}
Using (\ref{2ptlargex}) we can derive
the angle-dependent
correlation length $\xi(\theta)$ defined from the large-distance
behavior of the two-point function along a direction forming an angle $\theta$
with the side of the lattice.
Using the expression of $\xi(\theta)$ in terms of $M(\beta)$ 
reported, e.g., in \cite{MRR-85,CP-94}, we obtain\footnote{In particular, 
the correlation lengths along the side ($\theta=0$)
and the diagonal ($\theta=\pi/4$) of the lattice are
respectively given by 
$\xi_s^{-1} = - \ln {\rm tanh} \beta - 2\beta$ and
$\xi_d^{-1} = - \sqrt{2} \ln {\rm sinh} 2\beta$.}
\begin{equation}
\xi(\theta) = {1\over 4\beta_c u_t} \left[ 1 + 
\case{1}{6}\beta_c^2 \cos(4\theta) \,u_t^2 + O(u_t^4) \right].
\label{xi2}
\end{equation}
Thus, we see analytically that no correction of order $O(t^2)$ 
appears in the on-shell renormalization constant $Z(\beta)$---
both $T\bar T$ and ${\cal T}$ are absent. In $\xi(\theta)$
a $O(t^2)$ correction does appear as already observed in \cite{CPRV-98}. 
However, it is 
proportional to $\cos (4\theta)$, and thus it is due only 
to the leading operator breaking rotational invariance.
No contribution from the rotationally-invariant 
operator $T\bar T$ appears. 

The expansion of the Callan-Symanzik $\beta$-function can be derived 
using the same arguments employed by Nickel ~\cite{Nickel-82,Nickel-91}
in three dimensions.
Let us first consider the lattice Ising model and the 
coupling $g(t)$ defined in (\ref{defgl})
as a function of the reduced temperature. 
The expansion of $\chi$ and $\chi_4$ is well established:
\begin{eqnarray}
\chi &=& C_2 u_t^{-7/4} v_h^2 
   \left(1 + p_1 u_t^{7/4} + p_2 u_t^{11/4} \log u_t + p_3 u_t^{11/4} + \ldots 
   \right), \\
\chi_4 &=& C_4 u_t^{-11/2} v_h^4 
   \left(1 + p_4 u_t^{11/4} + \ldots \right),
\end{eqnarray}
where $C_2$, $C_4$, $p_1$, $p_2$, $p_3$, and $p_4$ are known constants
\cite{MW-73,GMC-88,McCoy-95,Nickel-99,BNNPSW-00,CHPV-00}. In particular 
$p_1 = -0.1081812\ldots$ Next, we determine the asymptotic behavior of 
$\mu_2 \equiv \sum_x x^2 \langle \sigma_0\sigma_x\rangle = 4 \chi \xi^2$ 
from its high-temperature expansion (HT).
The analysis of the 52nd-order HT expansion\footnote{
The HT expansion of $\mu_2$ can be found to $O(\beta^{36})$
in Ref.~\cite{Nickel-82}. The 52nd-order series has been kindly 
provided by Tony Guttmann \cite{Guttmann_private}.} of $\mu_2$
shows that its Wegner expansion can be written as
\begin{equation}
\mu_2 = A_2 u_t^{-15/4} v_h^2 
   \left(1 + p_5 u_t^2 + \ldots \right).
\label{eqmu2}
\end{equation}
The constant $p_5$ has been computed with high accuracy
in the following way. 
We have first defined a new series $s$ obtained by expanding 
in powers of $\beta$ the quantity 
$(\mu_2 u_t^{15/4} v_h^{-2}/A_2 - 1) u_t^{-2}$,
where $A_2 = 1.238136098$, and $u_t$, $v_h$ are given by Eqs.
(\ref{utdef}) and (\ref{vhdef}) truncated at order $t^3$ included.
Then, we analyzed $s$ by means of first-order inhomogeneous 
integral approximants biased to have a singularity at $\beta=\beta_c$. 
We verified that the critical exponent associated to the singularity 
is positive and then computed the value of $s$ for $\beta=\beta_c$.
We obtain finally\footnote{
It is worth noting that 
$p_5 \approx - \case{1}{2} C_2/A_2=-0.388722...$ within error bars,
so that $\mu_2$ can be written as
$\mu_2 = A_2 v_h^2 u_t^{-15/4}  -  \case{1}{2}\chi + \ldots $
This equation may be explained in terms of a momentum-dependence 
of the scaling field $u_h$. Indeed, $\mu_2$ is not a zero-momentum
quantity and thus it is related to the free energy in the 
presence of a nonuniform magnetic field $h(x)$. But, in this case
we expect additional contributions to the scaling fields, 
proportional to derivatives of $h(x)$ \cite{Wegner_1976}.
Our result for $\mu_2$ can be explained if the scaling field $u_h$ is 
a functional of $h(x)$ with a small-momentum behavior
$u_h = u_h|_{h={\rm const}} - \case{1}{8} \partial^2 h(x) + $
higher derivatives. Note also that the $u_t^2$ term in $\mu_2$ cannot be 
interpreted as a contribution due to irrelevant operators.
Indeed we do not expect $O(u_t^2)$ contributions 
associated with $T\bar{T}$, nor with 
the nonrotationally invariant $\cal T$, 
since $\mu_2$ is a rotationally invariant quantity. 
This point needs further investigation.
}
$p_5 = -0.388720(3)$.
It follows that 
\begin{equation}
g(t) = g^*\left[ 1 - p_1 u_t^{7/4} - p_5 u_t^2 + O(u_t^{11/4}\log u_t) \right].
\label{gt}
\end{equation}
Since the second-moment mass $m(t) = 1/\xi(t) = (4 \chi/\mu_2)^{1/2}$ scales 
as 
\begin{equation}
m(t)^2 = {4 C_2\over A_2} u_t^2 \left[ 1 + O(u_t^{7/4}) \right], 
\end{equation}
we obtain for the square-lattice Ising\footnote{
Notice that in the lattice model $g$ approaches $g^*$ from {\em above}
as $t\to 0$, while in the FT model the opposite happens.
For a discussion see \cite{LF-90} and references therein.}
$\beta$-function
\begin{equation}
\beta(g) \equiv m {\partial g \over \partial m}
= 2 m^2 \left( {d m^2\over du_t}\right)^{-1} {dg\over du_t} = 
 - {7\over 4} \Delta g
\left( 1 + b_1 |\Delta g|^{1/7}+ b_2 |\Delta g|^{2/7} + b_3 |\Delta g|^{3/7} 
+   \cdots \right)
\label{nan}
\end{equation}
where $\Delta g\equiv g^*-g$, and for the nonuniversal constant $b_1$, 
$b_1 = p_5 (-g^* p_1)^{-1/7}/(7 p_1) \approx 0.480(4)$.
It follows that $\beta'(g^*)=7/4$ and $p=1/7$.
Let us stress again that this value of $\beta'(g^*)$
is not related to the exponent of the leading irrelevant operator that
we expect to be two.
This phenomenon occurs whenever $\gamma < \omega_1\nu$. 
Indeed, in $g(t)$ there is a correction-to-scaling term proportional to
$t^\gamma$ because of the presence of an analytic background in the free energy
\cite{AF-83}. If $\gamma < \omega_1\nu$, it represents the 
leading nonanalytic correction in $g(t)$ and therefore 
$\beta'(g^*) = \gamma/\nu \not=\omega_1$.
It should be noted that such a phenomenon does not arise in three-dimensional
$O(N)$ models, where
the leading analytic corrections are determined by the leading irrelevant 
operator. For instance, for the three-dimensional Ising model
$\gamma/\nu =2-\eta \simeq 1.96 > \omega_1\simeq 0.8$.
We also mention the recent result $\beta'(g^*)\simeq 1.88$  
obtained in \cite{JS-99} using a numerical approach based on the 
high-temperature expansion of the Ising model, which is not too far from our 
exact prediction 7/4.

Now, the question is: which behavior should we expect for the $\phi^4$ field 
theory? In other words, does Eq. (\ref{gt}) holds for a generic model in the
$N=1$ $\phi^4$ universality class or are some terms absent?
And, in particular, are the conditions $p_1\not=0$ and $p_2\not=0$
a particular feature of the lattice Ising model only? 
Sometimes, see e.g. \cite{BB-85} and the discussion of \cite{PV-98},
it is conjectured that the $\beta$-function is analytic in 
FT models. However, it was shown in
\cite{PV-98} that this conjecture is not true: 
In the large-$N$ limit, nonanalytic terms are indeed present.
Unfortunately, in the two-dimensional case for 
$N=1$, we do not have any analytic control on the corrections to $\beta(g)$.
Nonetheless, we conjecture that 
Eq. (\ref{nan}) holds also for the FT $N=1$ model---of course, with different 
coefficients $b_1$, $b_2$ since the $\beta$-function is not universal.
We have essentially two arguments to support our conjecture:
\begin{description}
\item{a]} We do not see any reason why the bulk term that originates the 
$p_1 t^{7/4}$ contribution in (\ref{gt}) should not be present. 
Indeed, the analytic contribution is not a lattice artifact but has 
a well-defined FT meaning. In the CFT framework, it 
can be considered as a signature of the Identity operator and of its conformal
family. Thus, also for the FT model, we expect $p_1\not= 0$.
\item{b]} A $t^2$ correction is  certainly present in $g(t)$, since
we expect the operator $T\bar{T}$ to be present in the FT model. 
Thus $p_5$ will not be zero in (\ref{gt}), although it will be no longer
related to the correction appearing in $\mu_2$.
\end{description}
It is important to note that the strong nonanalytic corrections at $g=g^*$ 
we have found may explain the 
large observed deviations among the perturbative FT estimates of $g^*$ and 
$\beta'(g^*)$, the high-precision numerical results for $g^*$, 
and our prediction for $\beta'(g^*)$.
As a test, in App.~\ref{app2}
we have considered a simple Borel-summable function 
that has an asymptotic behavior of the form (\ref{nan}).
We have applied the standard resummation method presented above, observing 
large systematic deviations at $g=g^*$ and a systematic underestimate of 
the error bars.
We should note that these discrepances, although provide support
for the presence of strong nonanalytic corrections at $g=g^*$,
do not support our specific expansion (\ref{nan}). Indeed, even if $p_1=0$
in (\ref{gt}),
neglecting logarithmic terms, we would obtain
\be
\beta(g) = - 2 \Delta g\left(1 + c_1 |\Delta g|^{3/8} + \ldots\right).
\ee
Thus, also in this case, there would be a strong nonanalytic correction.

\section{$N\geq 3$ $\phi^4$ theory in $d=2$}

Let us now consider the multicomponent 
$\phi^4$ theory with $N\geq 3$. 
For $N=3$, the Pad\'e-Borel analysis of the five-loop series~\cite{OS-00} 
yields the
estimates $g^*=12.00(14)$ and $\beta'(g^*) = 1.33(2)$.
The result for $g^*$ is in reasonable agreement  with the 
more precise estimate $g^*=12.19(3)$ obtained by employing the
form-factor bootstrap approach~\cite{BNNPSW-99,BNNPSW-00}.
We shall now argue that the estimate $\beta'(g^*)\approx 4/3$ 
is again incorrect and that the correct value should instead be
$\beta'(g^*) = 2$.

The standard scenario predicts that, for $N\ge 3$, the theory is massive 
for all temperatures. The critical behavior is controlled by the 
zero-temperature Gaussian point and can be studied in perturbation theory
in the corresponding $N$-vector model. One finds only logarithmic
corrections to the purely Gaussian behavior.
It follows that the operators have dimensions that coincide 
with their naive (engineering)  dimensions, 
apart from logarithmic multiplicative
corrections related to the so-called anomalous dimensions. 
The leading irrelevant operator has dimension two \cite{BZLG-76b} and thus,
for $m\to 0$, we expect \cite{Symanzik-83}
\begin{equation}
g(m) = g^* \left\{ 1 + c \,m^2 \left( - \ln m^2\right)^\zeta 
\left[1 + O\left( {\ln (-\ln m^2)\over \ln m^2}\right)\right]\right\},
\label{gmsc}
\end{equation}
where $\zeta$ is an exponent related to the anomalous dimension of the
leading irrelevant operator, and $c$ is a constant. 
A one-loop calculation in the framework of the O($N$) $\sigma$ model
gives $\zeta=2/(N-2)$~\cite{BZLG-76b}.
Differentiating with respect to the mass, one obtains
\begin{equation}
\beta(g) = m {\partial g\over \partial m}= - 2 \Delta g 
   \left( 1 + {\zeta\over \ln \Delta g} + \cdots \right),
\label{bgn3}
\end{equation}
with $\Delta g\equiv g^*-g$.
Therefore, one expects $\beta'(g^*)=2$ with logarithmic corrections.

The expansion (\ref{bgn3}) for $\beta(g)$ is confirmed by a next-to-leading
order calculation in the framework of the large-$N$ 
expansion. Indeed, using the expression for $\beta(g)$ reported in 
\cite{CPRV-96} and
performing an asymptotic expansion around $g^*$ (see App. C for details), 
one finds
\begin{equation}
\beta(g) = 
- 2 (g^* - g) \left\{ 1 + {1\over N} \left[
  {2\over \ln \Theta} \left(1 + {l(\Theta)\over \ln \Theta}\right) +
  {5\over 2 \ln^2 \Theta} + O\left({l(\Theta)^2\over\ln^3 \Theta}\right)
 \right] \right\},
\label{bgnn}
\end{equation}
where $l(\Theta) \equiv \ln(-2 \ln \Theta)$ and $\Theta \equiv (g^* - g)/g^*$.
Comparing Eq.~(\ref{bgnn}) with Eq.~(\ref{bgn3}) we obtain
$\zeta=2/N+O(1/N^2)$, in agreement with the above-mentioned result 
$\zeta=2/(N -2)$.

Thus, for $N\ge 3$ we predict very strong nonanalytic corrections at 
$g=g^*$. A numerical study on a function with the asymptotic 
behavior (\ref{bgn3}) (see App. B) shows 
that such corrections give rise to a slow convergence of the perturbative 
resummations. In particular, the estimate of $\omega_1$ may 
be incorrect in spite of the stability of the results with 
the number of loops considered in the analysis. It is thus not 
surprising that Ref. \cite{OS-00} find $\beta'(g^*)\simeq 4/3$ 
instead of the correct result $\beta'(g^*) = 2$.

\section*{Acknowledgements}

We thank Tony Guttmann 
for making available to us the 52nd-order HT series
of the second moment for the Ising model
and for informing us of Ref. \cite{Guttmann-2000}.
Moreover, we thank
Martin Hasenbusch for discussions,
Alan Sokal for a critical reading of a first draft of the manuscript, 
and Alexander Sokolov for sending us Ref. \cite{OS-00}.
This work was partially supported by the 
European Commission TMR programme ERBFMRX-CT96-0045.

\appendix

\section{Nonunitary extension of the Ising model} \label{app1}

\renewcommand{\theequation}{A.\arabic{equation}}
\setcounter{equation}{0}

The 4/3 operator appears in a nonunitary extension of the Ising model that
describes Ising percolation.

Let us first of all explain what we mean with the notion of ``nonunitary
extension'' of the Ising universality class. The starting point is the
classification of the minimal unitary conformal field theories discussed
in~\cite{FQS-84,BPZ-84}.

The operator content of the unitary  CFT's {\sl that only possess a $Z_2$
symmetry} (like the $\phi^4$ theory and its multicritical generalizations)
is defined by the weights:
\eq
h_{p,q}=\frac{[(m+1)p-mq]^2-1}{4m(m+1)}
\label{kac}
\en
with $m=3,4,5\cdots$ and the constraints $1\leq p\leq (m-1)$,
  $1\leq q\leq p$. The relation between $h$ and the renormalization-group eigenvalue $y$ is
$y=2-2h$. For the Ising model $m=3$. Higher values of $m$ correspond to
multicritical Ising-like
models (i.e.  theories with a $Z_2$ symmetric potential with powers up to
 $\phi^{2m-2}$). These are the continuum-limit CFT's that correspond to the
 models introduced in~\cite{ABF-84,Huse-84}. 
With $m=3$ we have only three allowed combinations of $(p,q)$:
(1,1), (2,1) and (2,2) that correspond to the identity, energy and spin
operators of the Ising model. They are called ``primary'' fields. 
{}From any one of these primary fields one has then an
infinite tower of  ``secondary'' fields whose scaling dimensions are shifted by
integers with respect to those of the primary fields. Since in the Ising model
{\sl all the primary fields are relevant}, all the irrelevant fields must be
shifted by integers, hence they cannot be distinguished from the analytic
corrections. This is the only model in which this happens. In all other
models with $m>3$, there are primary fields that are irrelevant and hence
are candidates for nontrivial subleading scaling dimensions.

Besides unitary theories, there is an infinite set of nonunitary ones 
for all the rational (but noninteger) values of $m$. Apart from the fact that
they do not fulfill unitarity, they have the same properties of those with
integer $m$. In particular, 
their operator content is completely known and closed
expressions for the correlators exist.
 These models (with both integer and non integer values of $m$)
 are usually called  Rational Conformal Field
 Theories (RCFT).

However this is not the end of the story. In the last few years it has been
realized that it is possible to give a meaning, in the framework of the so
called Logarithmic Conformal Field Theories (LCFT)~\cite{g-93},
 also to more general 
theories, obtained by including in the operator algebra some of
 the operators  corresponding to the
 values of $p$ and $q$ excluded in eq.~(\ref{kac})~\cite{f-96}.

For instance, in the Ising case (i.e. $m=3$) in which we are interested 
 one should enlarge the set of operators of the standard Ising CFT to those 
 of the type
 $h_{3,n}$, $n=1,2,\cdots$ and 
$h_{k,4}$, with $k=1,2\cdots$. The  LCFT 
 obtained in this way is what we mean by 
 ``nonunitary extension'' of the Ising model.

Despite the fact that these theories are much more difficult to study than the
 standard RCFT's, 
several interesting results have been obtained in these last years
(for a recent account see for instance~\cite{f-97,gk-99} and references
therein). For the purpose of the present paper we only need to know the scaling
dimensions of the new operators. These can be easily obtained by looking at
eq.~(\ref{kac}).

In particular, in the Ising case, we see that $h_{3,1}=\frac53$ hence 
$y_{3,1}=-\frac43$, which is exactly the irrelevant operator that we are looking
for. Further examples 
of such operators (only the relevant ones are listed)
are:

$h_{3,2}=\frac{35}{48}$ hence 
$y_{3,2}=\frac{13}{24}$,

$h_{3,3}=\frac{1}{6}$ hence 
$y_{3,3}=\frac{5}{3}$,

$h_{2,4}=\frac{5}{16}$ hence 
$y_{2,4}=\frac{11}{8}$.

Note that, for
all the values  $m>3$ (i.e in the multicritical models), 
Eq.~(\ref{kac}) admits a 
 unitary, well defined, operator of type $h_{3,1}$  with
weight $(m+2)/m$ so that $y=-4/m$. Thus, a naive
limit $m\to 3$ would lead to an operator with $y=-4/3$,
This argument is usually given to support the existence of a 
scaling operator with $\omega_1=4/3$ (see e.g. \cite{ZJ-book}). 
However, as we have seen, exactly
for $m=3$ this operator becomes ``border-line'' and 
it does not belong anymore to the Ising universality class, but only to its
nonunitary extension. Thus, the limit $m\to 3$ of Eq.~(\ref{kac}) cannot
be considered as an indication in favor of the presence of
a $\omega_1=4/3$ field in the (unitary) Ising universality class,
which the $\phi^4$ theory belongs to.

Another context in which the $y=4/3$ field appears, which is completely 
independent and allows to share some more light on its meaning, is the
Coulomb gas approach to the $q$-state Potts models  due
to Nienhuis~\cite{Nienhuis-82}. By mapping the Potts model in a suitable 
vertex-type model Nienhuis was able to identify both the leading and the 
subleading thermal and magnetic operators as a function of $q$.
 For $q=2$ the subleading thermal operator is exactly
 $y=-4/3$ and the subleading magnetic operator is 
$y=13/24$ (see also~\cite{N87}).
However, as already noted in~\cite{Nienhuis-82},
{\sl these are operators of the Vertex model and not of the Ising model} and
they decouple for $q=2$. In other words, the Vertex model of Nienhuis
is a good candidate for an exactly solvable model
whose continuum limit is the nonunitary extension of the Ising model. If one
requires the Vertex model to have a ``physical spectrum'' according to the
definition given in~\cite{Nienhuis-82}, then one selects only the operators of
the standard Ising
model and the $y=4/3$ operator decouples. The requirement of having
a ``physical spectrum'' is equivalent to impose unitarity on the model.

It would be nice to have some kind of insight of the physical meaning of the
above-mentioned operators directly from Ising model. Some hints in this 
direction are
given by the so called ``Ising percolation'' problem
i.e. the  behaviour of the Coniglio-Klein clusters in the Ising model.
It turns out that the {\sl relevant} operators in the nonunitary extension of
the Ising universality class (i.e. both the standard ones $y=1$ and $y=15/8$
and the
``border-line'' ones $y=13/24$, $y=11/8$ and $y=5/3$) become fractal dimensions 
of suitable
sets of links (or sites) of the Ising percolation model at the critical point.
In particular, $y=15/8$ is the fractal dimension of the percolating cluster,
$y=1$ is related to the correlation length, $y=13/24$ is the fractal dimension
of the red bonds (see~\cite{sv1}), $y=5/3$ is
the fractal dimension of the percolating cluster in the presence of a boundary
(see~\cite{sv2}), and $y=11/8$  is the fractal dimension of the hull
(see~\cite{cn86}). Unfortunately, the operator in which we are interested, being
irrelevant, cannot
be realized  as a fractal dimension, but the coincidence of the other indices
strongly supports the idea that it  
should also appear as subleading dimension of some
suitably chosen set of links. 

Some theoretical justification of this remarkable
coincidence of critical indices and fractal dimensions 
can be found in an interesting conjecture that was
proposed for the first time in~\cite{th84} and then discussed in detail
in~\cite{sv1}  and~\cite{sv2}.
According to this conjecture, Ising percolation  is described 
by the $q\to 1$ limit of the tricritical $q$-state Potts model in exactly the 
same way in which the $q\to 1$ limit of ordinary $q$-state Potts describes 
standard percolation. The operator content of the $q\to 1$ limit of the 
tricritical $q$-state Potts model can be studied with the same Coulomb gas
techniques discussed above. It turns out that it contains (together with other
operators) the nonunitary extension of the Ising model, and thus explains the
above coincidence of critical indices and fractal dimensions. Notice that this
conjecture is further supported by the identification as fractal dimensions of
suitable sets of links of other critical indices that belong to the 
$q\to 1$ limit of the tricritical Potts but that are outside
the nonunitary Ising class---see~\cite{sv2} for a discussion.

\section{Resummation of simple test functions} \label{app2}
\renewcommand{\theequation}{B.\arabic{equation}}
\setcounter{equation}{0}

In this appendix 
we consider a simple test function which behaves as (\ref{bgexp})
and whose perturbative expansion around $g=0$ is divergent but Borel summable.
We show 
that many terms are needed in order to obtain the correct results, and, 
even worse, that in this case the standard method to set the error bars 
does not work properly. The estimated errors are much smaller than
the difference between the estimate and the exact value.

Consider the function
\begin{equation}
Z(g) = {1\over \sqrt{2\pi}} \int_{-\infty}^{+\infty} dx\, 
\exp \left( - \case{1}{2}x^2 - \case{1}{4!} g x^4 \right).
\end{equation}
Its expansion in powers of $g$, $Z(g) = \sum_k Z_k g^k$, 
is Borel summable, 
and the large-order behavior of the $k$th-order coefficient $Z_k$ is given by   
\begin{equation}
Z_k =  (-1)^k {(4k-1)!!\over 4!^k k!} \propto \left( -{2\over 3}\right)^k (k-1)!\left[ 1 + O(1/k)\right].  
\end{equation} 
The function
$Z(g)$ is analytic in the complex plane with a cut along the negative real
axis, and in particular it is analytic for $g=1$.
For $\delta\equiv 1 - g\to 0$
it behaves as
\begin{equation}
Z(g) = Z_0 + Z_1 \delta + O(\delta^2),
\label{zbh}
\end{equation}
where 
$Z_0= 0.9189189...$ and $Z_1= -0.0573155...$
In this case, in which the function is analytic,
the resummation method we presented in the Introduction 
provides good estimates of the constants appearing in (\ref{zbh}).
One indeed obtains 
$Z_0 = 0.9189(1)$ and  $Z_1 = -0.0572(3) $ from the 5th-order series, and
$Z_0 = 0.918919(1)$ and  $Z_1 = -0.057315(3)$ from the 10th-order 
series.\footnote{
The estimates and their errors are obtained using the 
procedure of \cite{PV-98}.
The estimate is obtained from the ``optimal'' values of the two
free parameters introduced in the procedure ($b$ and $\alpha$),
which are determined  by maximizing the stability of the results
with respect to the order of the series analyzed.
The errors are related to the stability 
of the results with respect to variations of the free parameters $b$ and 
$\alpha$ around their optimal values.} Most important, the method 
provides correct estimates of the errors.

In order to reproduce a nonanalytic behavior
similar to (\ref{nan}), we consider the function
\begin{equation}
B(b,g) = Z(g) + c (1-g)^{1+b}.
\end{equation}
Setting $c=Z_1$, we have for $g\to 1$
\begin{equation}
B(b,g) = Z_0 + Z_1 \delta \left( 1 + \delta^{b} \right) +  O(\delta^2).
\end{equation}
We apply the same resummation procedure used for $Z(g)$ 
to the perturbative expansion of $B(b,g)$.
To reproduce the correction predicted in the Ising case, we fix $b=1/7$.
The results of the analysis are now much less satisfactory. Indeed, we find
$Z_0 = 0.916(6)$ and $Z_1=-0.112(2)$ from the 5th-order series,
and $Z_0 = 0.918(1)$ and $Z_1=-0.103(6)$ from the 10th-order series.
The estimate of $Z_0$ is not as precise as before, but the error
is still correct. This is not surprising since the nonanalyticity is here 
rather weak, the nonanalytic corrections being of order $\delta^{1+b}$. 
On the other hand, the estimate of $Z_1$, which is determined
by resumming the $dB(b,g)/dg$ (here, the nonanalytic corrections are 
stronger, of order $\delta^b$), is very imprecise and the 
estimate of the error, which is  obtained from the stability analysis,  
is completely incorrect: the five-loop estimate differs from the exact value
by more than 25 estimated error bars! Moreover, 
extending the series appears to be 
of little help. We conjecture that a similar phenomenon is happening 
in the FT estimates for $N=1$. Although the perturbative results
indicate $\omega_1\approx 4/3$ with a tiny error, the correct result 
is sensibly different.

We have also considered the case in which we add a term of the 
form $Z_1 g/\log(1 - g)$, 
which mimicks the behaviour of the $\beta$-function for $N\ge 3$,
observing completely analogous deviations.

We have repeated the exercise by considering a nonanalytic singularity
similar to that expected in three dimensions, i.e. by setting $b\simeq 1$.
For example, for $b=9/10$ we find 
$Z_0 = 0.917(1)$ and $Z_1=-0.068(4)$ from the 5th-order series,
$Z_0 = 0.9186(1)$ and $Z_1=-0.060(2)$ from the 10th-order series.
%for $b=11/10$ we found 
%$Z(1) = 0.9178(6)$ and $Z'(1)=-0.054(3)$ from the 5th-order series,
%and $Z(1) = 0.91916(5)$ and $Z'(1)=-0.0548(2)$ from the 10th-order series.
As expected, the effect of the nonanalyticity is much smaller and the errors
reasonable, although slightly underestimated.

\section{Asymptotic expansion of large-$N$ integrals} \label{app3}
\renewcommand{\theequation}{C.\arabic{equation}}
\setcounter{equation}{0}

In this Appendix we wish to compute the asymptotic expansion for 
$\Theta\to 0$ of integrals of the form 
\be
I_n(f,\Theta) = \int_0^\infty du\, {f(u)\over [\Theta + \delta(u)]^n},
\ee
where $f(u)\sim u^{-p}$ for $u\to\infty$, and $n$ and $p$ are integers
satisfying $n\ge1$, $p\ge2$. The function $\delta(u)$ is given by
\be
\delta(u) = - {2\over u\xi} \log{1 - \xi\over 1 + \xi},
\ee
where
\be
\xi(u) = \sqrt{u\over u + 4}.
\ee
The results presented here extend App. A of \cite{PV-98} to two dimensions.
We wish to compute the leading nonanalytic contributions to the 
asymptotic expansion. For this purpose, we can replace $\delta(u)$ and $f(u)$
with their leading behaviour for $u\to\infty$ and write
\be
I_n(f,\Theta) \approx \int_{1/\Lambda}^\infty 
  {du\, u^{-p}\over [\Theta + (2 \log u)/u]^n},
\label{eq:1}
\ee
where $\Lambda$ is an arbitrary cutoff satisfying $0 <  \Lambda <1$. 
Then we make the substitution 
\be
 {2\over u} \log u = y.
\label{substitution}
\ee
For $y\to 0$, $u\to\infty$, Eq. (\ref{substitution}) can be solved, obtaining
the asymptotic expansion 
\be
{1\over u} = - {y\over 2\log (y/2)} 
 \left\{1 + \sum_{n=1}^\infty\sum_{m=1}^n a_{nm} 
  {[\log(-\log (y/2))]^m\over \log^n (y/2)} \right\}.
\ee
The first coefficients are: $a_{11}=a_{22}=1$, $a_{21}=-1$.

Substituting this expression in (\ref{eq:1}) and keeping only the 
leading contributions, we obtain
\be
I_n(f,\Theta) \approx \int_0^\Lambda {dy\over y u^{p-1}} 
    {1\over (\Theta + y)^n},
\ee
where analytic terms have been systematically neglected.

Since $p\ge 2$, we see that $I_n(f,\Theta)$ can be written as a sum 
of terms of the form
\be
K_{nmp}(\Theta) = \int_0^\Lambda dy\,
  {[\log(-\log y)]^p\over (-\log y)^m (\Theta + y)^n},
\ee
with $m$, $n$, and $p$ integers.
The nonanalytic terms are due to the integrals with $n\ge 1$, and thus we 
consider only this case.
Now, observe that we need to consider 
$n=1$ only , since
\be
K_{nmp}(\Theta) = {(-1)^{n-1}\over (n-1)!} {d^{n-1}\over d\Theta^{n-1}}
   K_{1mp}(\Theta).
\ee
Then, note also that 
\be
[\log(-\log y)]^p = \lim_{\epsilon\to 0} 
\left[{ (-\log y)^\epsilon - 1 \over \epsilon} \right]^p.
\ee
Thus, it is enough to consider $K_{1\alpha 0}$, where $\alpha$ is a real
number.  In the following we assume $\alpha > 1$. The final result
however will be correct for all values of $\alpha$.
To compute the asymptotic expansion, first perform a Mellin transformation,
rewriting
\be
K_{1\alpha 0} = - \int_{-1/2-i\infty}^{-1/2+i\infty} 
  {ds\over 2\pi i} \, {\pi\over \sin \pi s}\ \Theta^s \ R_\alpha(\Lambda,s),
\ee
where
\be
R_\alpha(\Lambda,s) = \int_0^\Lambda {dy\over (-\log y)^\alpha} y^{-1-s} =\, 
       \int_{-\log \Lambda}^\infty {dt\over t^\alpha} e^{st}.
\label{eq:2}
\ee
The previous equation defines $R_\alpha(\Lambda,s)$ for ${\rm Re}\ s \le 0$.
By rotating the $t$ contour one can obtain an analytic continuation
in the domain ${\rm Re}\ s > 0$ with a cut along the positive real axis.
In the following, we need the discontinuity at the cut. A simple 
calculation gives
\be
R_\alpha(\Lambda,s+) - R_\alpha(\Lambda,s-) = 
\int_C {dt\over t^\alpha} e^{st} = {2 \pi i\over \Gamma(\alpha)}\ s^{\alpha-1},
\ee
where $C$ is a contour running counterclockwise around the negative $t$-axis.
We also need $R_\alpha(\Lambda,0) = (- \log \Lambda)^{1-\alpha}/(\alpha -1)$.
In order to compute the asymptotic expansion of 
$K_{1\alpha 0}(\Theta)$, deform the $s$-integral, so that it goes
around the positive $s$-axis. Keeping into account the pole at $s=0$ we obtain
\be
K_{1\alpha 0}(\Theta) = R_\alpha(\Lambda,0) -  
    \int_0^\mu {ds\over 2\pi i} {\pi\over \sin \pi s} \Theta^s 
   [R_\alpha(\Lambda,s+) - R_\alpha(\Lambda,s-)] -
  \int_{C_+ + C_-} {ds\over 2\pi i} {\pi\over \sin \pi s} \Theta^s
     R_\alpha(\Lambda,s),
\ee
where $0<\mu<1$ is arbitrary and $C_\pm = \{s:\ {\rm Re}\ s = \mu,\;\;
   \pm{\rm Im} s > 0\}$. The integral over the lines $C_\pm$ is of order 
$\Theta^\mu$ and can therefore be discarded. In order to compute the 
integral over the cut,
we make the substitution $- s \log \Theta = t$, expand
the integrand in powers of $1/\log \Theta$ and replace the upper integration 
limit $- \mu \log \Theta$ with $\infty$ ---again we make an error of 
order $\Theta^\mu$. The final integrations are trivial.
We obtain finally
\be
K_{1\alpha 0}(\Theta) \approx {1\over \alpha-1} (-\log \Lambda)^{1-\alpha} -
   (-\log \Theta)^{1-\alpha} \sum_{k=0}^\infty b_k 
   {\Gamma(2k+\alpha-1)\over \Gamma(\alpha)} 
   \left({\pi\over \log \Theta}\right)^{2k},
\ee
where the coefficients $b_k$ are defined by
\be
{1\over \sin x} = \, \sum_{k=0}^\infty b_k x^{2k-1}.
\ee


\begin{thebibliography}{199}

\bibitem{Parisi-80} G.~Parisi, Carg\`{e}se Lectures (1973),
J.\ Stat.\ Phys.\ {\bf 23} (1980) 49.
%%CITATION = JSTPB,23,49;%%

\bibitem{BNGM-77} G.~A.~Baker, Jr.,  B.~G.~Nickel, 
M.~S.~Green, and D.~I.~Meiron,
Phys.\ Rev.\ Lett.\ {\bf 36} (1977) 1351;
G.~A.~Baker, Jr., B.~G.~Nickel, and D.~I.~Meiron,
Phys.\ Rev.\ B {\bf 17} (1978) 1365.  
%%CITATION = PRLTA,36,1351;%%
%%CITATION = PHRVA,B17,1365;%%

\bibitem{OS-00}
E.~V.~Orlov and A.~I.~Sokolov,
``Critical thermodinamics of the two-dimensional systems in
five-loop renormalization-group approximation,'' (in Russian),
to appear in Fiz. Tverd. Tela (2000). 
A shorter English version appears as e-print hep-th/0003140.
%%CITATION = HEP-TH 0003140;%%

\bibitem{ZJ-book}  J.~Zinn-Justin,
{\em Quantum Field Theory and Critical Phenomena},
third edition (Clarendon Press, Oxford, 1996).

\bibitem{Lipatov-77} 
L.~N.~Lipatov, 
Zh. Eksp. Teor. Fiz. {\bf 72} (1977) 411 
[Sov. Phys. JETP {\bf 45} (1977) 216].
%%CITATION = ZETFA,72,411;%%
%%CITATION = SPHJA,45,216;%%

\bibitem{BLZ-77}
E.~Br\'ezin, J.~C.~Le Guillou, and J.~Zinn-Justin,
Phys. Rev. {\bf D 15} (1977) 1544, 1588.
%%CITATION = PHRVA,D15,1544;%%
%%CITATION = PHRVA,D15,1588;%%

\bibitem{LZ-77} J.~C.~Le Guillou and J.~Zinn-Justin,
Phys.\ Rev.\ Lett.\ {\bf 39} (1977) 95; 
Phys.\ Rev.\ B {\bf 21} (1980) 3976.
%%CITATION = PRLTA,39,95;%%
%%CITATION = PHRVA,B21,3976;%%

\bibitem{Nickel-82}
B.~G.~Nickel, in {\em Phase Transitions},
M.~L\'evy, J.~C.~Le~Guillou, and J.~Zinn-Justin eds.,
(Plenum, New York and London, 1982), p. 291.
%%SLAC
% @Article{Nickel:1980he,
%     author    = "B. G. Nickel",
%     title     = "THE PROBLEM OF CONFLUENT SINGULARITIES",
%     note     = "In *Cargese 1980, Proceedings, Phase Transitions*, 291-324"
% }

\bibitem{Nickel-91} 
B.~G.~Nickel,  Physica A {\bf 117} (1991) 189.
%%CITATION = PHYSA,A177,189;%%

\bibitem{Sokal-94} 
A. D. Sokal, Europhys. Lett. {\bf 27} (1994) 661; (E) {\bf 30} (1995) 123;
B. Li, N. Madras, and A. D. Sokal, J. Stat. Phys. {\bf 80} (1995) 661.
%%CITATION = JSTPB,80,661;%%
%%CITATION = EULEE,27,661;%%
%%CITATION = EULEE,30,123;%%

\bibitem{BB-97}
C. Bagnuls and C. Bervillier, J. Phys. Stud. {\bf 1} (1997) 366.

\bibitem{PV-98} 
A.~Pelissetto and E.~Vicari, Nucl. Phys. B {\bf 519} (1998) 626;
Nucl. \ Phys.\ B (Proc. Suppl.) {\bf 73} (1999) 775;
Nucl. \ Phys.\ {\bf B575} (2000) 579.
%%CITATION = NUPHZ,73,775;%%
%%CITATION = COND-MAT 9911452;%%

\bibitem{CPRV-96}
M.~Campostrini, A.~Pelissetto,
P.~Rossi, and E.~Vicari, Nucl.\ Phys.\ B {\bf 459} (1996) 207.
%%CITATION = NUPHA,459,207;%%

\bibitem{GZ-98} R.~Guida and J.~Zinn-Justin, 
J.~Phys. A {\bf 31} (1998) 8103.
%%CITATION = JPAGB,31,8103;%%

\bibitem{CPRV-99}
M. Campostrini, A. Pelissetto, P. Rossi, and E. Vicari,
Phys. Rev. E {\bf 60} (1999) 3526.
%%CITATION = PHRVA,E60,3526;%%

\bibitem{Hasenbusch-99}
M. Hasenbusch, J. Phys. A {\bf 32} (1999) 4851.
%%CITATION = JPAGB,32,4851;%%

\bibitem{CPRV-00}
M.~Campostrini, A.~Pelissetto, P.~Rossi, and E.~Vicari,
Phys. Rev. B {\bf 61} (2000) 5905;
%% ``The critical equation of state of three-dimensional XY systems,''
%% e-print cond-mat/0001440, Phys. Rev. B {\bf 62} (2000), in press.
Phys. Rev. B {\bf 62} (2000) 5843.
%%CITATION = PHRVA,B61,5905;%%
%%CITATION = COND-MAT 0001440;%%

\bibitem{HT-99}
M. Hasenbusch and T. T\"or\"ok, J. Phys. A {\bf 32} (1999) 6361.
%%CITATION = JPAGB,32,6361;%%

\bibitem{Zinn-Justin-00}
J. Zinn-Justin,
``Precise determination of critical exponents and 
equation of state by field theory methods,''
e-print hep-th/0002136.
%%CITATION = HEP-TH 0002136;%%

\bibitem{GRNR-76}
G.~R.~Golner and E.~K.~Riedel,
Phys.\ Lett.\ A {\bf 58} (1976) 11;
K.~E.~Newman and E.~K.~Riedel,
Phys.\ Rev.\ B {\bf 30} (1984) 6615.
%%CITATION = PHLTA,A58,11;%%
%%CITATION = PHRVA,B30,6615;%%

\bibitem{CHPV-00}
M. Caselle, M. Hasenbusch, A. Pelissetto, and E. Vicari,
``High-precision estimate of $g_4$ in the 2D Ising model,''
e-print hep-th/0003049.
%%CITATION = HEP-TH 0003049;%%

\bibitem{BNNPSW-00} J.~Balog, M.~Niedermaier, F.~Niedermayer, A.~Patrascioiu, 
E.~Seiler, and P.~Weisz, 
%% ``The intrinsic coupling in integrable quantum field theories,''
%% e-print  hep-th/0001097.
Nucl. Phys. B {\bf 583} (2000) 614.
%%CITATION = HEP-TH 0001097;%%

\bibitem{BC-96} 
P. Butera and M. Comi, Phys. Rev. B {\bf 54} (1996) 15828.

\bibitem{ZLF-96}
S. Zinn, S.-N. Lai, and M. E. Fisher,
Phys. Rev. E {\bf 54} (1996) 1176.

\bibitem{MW-73} B. M.~McCoy and T. T.~Wu, {\sl The two dimensional Ising Model}
(Harvard Univ. Press, Cambridge, 1973).

\bibitem{WMTB-76}
T. T. Wu, B. M. McCoy, C. A. Tracy, and E. Barouch,
Phys. Rev. B {\bf 13} (1976) 316.
%%CITATION = PHRVA,B13,316;%%

\bibitem{McCoy-95}
B. M.~McCoy, in 
{\sl Statistical Mechanics and Field Theory}, eds. V. V. Bazhanov
and C. J. Burden (World Scientific, Singapore, 1995).

\bibitem{FQS-84}
D.~Friedan, Z. Qiu, and S. Shenker,
Phys. Rev. Lett. {\bf 52} (1984) 1575.
%%CITATION = PRLTA,52,1575;%%

\bibitem{BPZ-84}
A.~A.~Belavin, A.~M.~Polyakov and A.~B.~Zamolodchikov,
Nucl. Phys. B {\bf 241} (1984) 333.
%%CITATION = NUPHA,B241,333;%%

\bibitem{BF-85}
M.~Barma and M. E. Fisher,
Phys. Rev. B {\bf 31} (1985) 5954.
%%CITATION = PHRVA,B31,5954;%%

\bibitem{CH-00} M.~Caselle and M.~Hasenbusch, 
%% ``Critical amplitudes and mass spectrum of the 2D Ising model in
%% a magnetic field,''
%% e-print hep-th/9911216, Nucl. Phys. B to appear.
Nucl. Phys. B {\bf 579} (2000) 667.
%%CITATION = HEP-TH 9911216;%%

\bibitem{AF-83} 
A. Aharony and M. E. Fisher, Phys. Rev. B {\bf 27} (1983) 4394.
%%CITATION = PHRVA,B27,4394;%%

\bibitem{Nickel-99}
B. Nickel, J. Phys. A {\bf 32} (1999) 3889; 
{\bf 33} (2000) 1693.
%%CITATION = JPAGB,32,3889;%%
%%CITATION = JPAGB,33,1693;%%

\bibitem{Guttmann-2000} 
W. P. Orrick, B. Nickel, A. J. Guttmann, and J. H. H. Perk,
``The susceptibility of the square lattice Ising model: New developments,"
to be submitted to J. Stat. Phys.

\bibitem{Queiroz-00} S. L. A. de Queiroz, J. Phys. A {\bf 33} (2000) 721.
%cond-mat/9912090.
%%CITATION = JPAGB,33,721;%%

\bibitem{Hasenbusch-2000} 
M. Hasenbusch,  in preparation.

\bibitem{CW-67}
H. Cheng and T. T. Wu,
Phys. Rev. {\bf 164} (1967) 719.
%%CITATION = PHRVA,164,719;%%

\bibitem{SS-99} J. Salas and A. D. Sokal, 
``Universal amplitude ratios in the critical two-dimensional 
Ising model on a torus,''
e-print cond-mat/9904038v1; 
J. Stat. Phys. {\bf 98} (2000) 551 [cond-mat/9904038v2].
%%CITATION = COND-MAT 9904038;%%

\bibitem{MRR-85}
V. F. M\"uller, T. Raddatz, and W. R\"uhl,
Nucl. Phys. B {\bf 251} [FS13] (1985) 212; (E)
Nucl. Phys. B {\bf 259} (1985) 745.
%%CITATION = NUPHA,251,212;%%
%%CITATION = NUPHA,259,745;%%

\bibitem{CP-94}
S. Caracciolo and A. Pelissetto,
Nucl. Phys. B {\bf 420} (1994) 141.
%%CITATION = NUPHA,420,141;%%

\bibitem{CPRV-98}
M. Campostrini, A. Pelissetto, P. Rossi, and E. Vicari,
Europhys.\ Lett.\ {\bf 38} (1997) 577;
Phys.\ Rev.\ {\bf E 57} (1998) 184;
Nucl. Phys. B (Proc. Suppl.) {\bf 53} (1997) 690.
%%CITATION = EULEE,38,577;%%
%%CITATION = PHRVA,E57,184;%%
%%CITATION = NUPHZ,53,690;%%

\bibitem{GMC-88}
S. Gartenhaus and W. S. McCullough, Phys. Rev. {\bf B} 38 (1988) 11688.
%%CITATION = PHRVA,B38,11688;%%

\bibitem{Guttmann_private}
A. J. Guttmann, private communication, 2000.

\bibitem{Wegner_1976}
F. J. Wegner, in {\sl Phase transitions and critical phenomena, Vol. 6}
eds. C. Domb and M. Green (New York, Academic Press, 1976), p. 7.

\bibitem{LF-90}
A. J. Liu and M. E. Fisher, Physica A {\bf 156} (1989) 35.

\bibitem{JS-99}
G. Jug and B. N. Shalaev, J. Phys. A {\bf 32} (1999) 7249.
%%CITATION = JPAGB,32,7249;%%

\bibitem{BB-85} 
C. Bagnuls and C. Bervillier, Phys. Rev. B {\bf 32} (1985) 7209.
%%CITATION = PHRVA,B32,7209;%%

\bibitem{BNNPSW-99} 
J.~Balog, M.~Niedermaier, F.~Niedermayer, A.~Patrascioiu, 
E.~Seiler, and P.~Weisz, Phys. Rev. D {\bf 60} (1999) 094508.
%%CITATION = PHRVA,D60,094508;%%

\bibitem{BZLG-76b}
E. Br\'ezin, J. Zinn-Justin, and J. C. Le Guillou,
Phys. Rev. B {\bf 14} (1976) 4976.
%%CITATION = PHRVA,B14,4976;%%

\bibitem{Symanzik-83}
K. Symanzik, Nucl. Phys. B {\bf 226} (1983) 187, 205.
%%CITATION = NUPHA,226,187;%%
%%CITATION = NUPHA,226,205;%%

\bibitem{ABF-84}
G.~E.~Andrews, R.~J.~Baxter, and P.~J.~Forrester,
J. Stat. Phys. {\bf 35} (1984) 193.
%%CITATION = JSTPB,35,193;%%

\bibitem{Huse-84}
D.~A.~Huse, Phys. Rev. B {\bf 30} (1984) 3908.
%%CITATION = PHRVA,B30,3908;%%

\bibitem{g-93}
V.~Gurarie,
%``Logarithmic operators in conformal field theory,''
Nucl.\ Phys.\  {\bf B410} (1993) 535.
%%CITATION = HEP-TH 9303160;%%

\bibitem{f-96}
M.~A.~Flohr,
%``On Modular Invariant Partition Functions of Conformal Field Theories with Logarithmic Operators,''
Int.\ J.\ Mod.\ Phys.\  {\bf A11} (1996) 4147.
%%CITATION = HEP-TH 9509166;%%

M.~A.~Flohr,
%``On Fusion Rules in Logarithmic Conformal Field Theories,''
Int.\ J.\ Mod.\ Phys.\  {\bf A12} (1997) 1943.
%%CITATION = HEP-TH 9605151;%%

\bibitem{f-97}
M.~A.~Flohr,
%``Singular vectors in logarithmic conformal field theories,''
Nucl.\ Phys.\  {\bf B514} (1998) 523.
%%CITATION = HEP-TH 9707090;%%

\bibitem{gk-99}
M.~R.~Gaberdiel and H.~G.~Kausch,
%``A local logarithmic conformal field theory,''
Nucl.\ Phys.\  {\bf B538} (1999) 631.
%%CITATION = HEP-TH 9807091;%%

\bibitem{Nienhuis-82}
B. Nienhuis, J. Phys. A {\bf 15} (1982) 199.
%%CITATION = JPAGB,15,199;%%

\bibitem{N87}
B. Nienhuis, in {\sl Phase transitions and critical phenomena, Vol. 11}
eds. C. Domb and J. L. Lebowitz (New York, Academic Press), p. 1.

\bibitem{sv1}
A. L. Stella and C. Vanderzande,
\PRL{62} (1989) 1067.
%%CITATION = PRLTA,62,1067;%%

\bibitem{sv2}
A. L. Stella and C. Vanderzande,
J. Phys. A {\bf 22} (1989) L445.
%%CITATION = JPAGB,22,L445;%%

\bibitem{cn86}
J. L. Cambier and M. Nauenberg,
Phys. Rev. B {\bf 34} (1986) 8071.
%%CITATION = PHRVA,B34,8071;%%

\bibitem{th84}
T. Temesv\'ari and L. Her\'enyi
J. Phys. A {\bf 17} (1984) 1703.
%%CITATION = JPAGB,17,1703;%%


\end{thebibliography}
\end{document}